\documentstyle[12pt,aaspp4]{article}

\def\ha{H$\alpha$}
\def\o3{[\ion{O}{3}]}
\def\s2{[\ion{S}{2}]}
\def\kms{\ifmmode{{\rm \,km\,s^{-1}}}\else{${\rm \,km\,s^{-1}}$}\fi}
\def\hm#1#2{\ifmmode{ #1^{\rm h} #2^{\rm m}}\else{ $ #1^{\rm h} #2^{\rm m}$}\fi}
\def\hms#1#2#3{\ifmmode{#1^{\rm h} #2^{\rm m} #3^{\rm s}} 
\else{$ #1^{\rm h} #2^{\rm m} #3^{\rm s}$}\fi}

\lefthead{Levenson et al.}
\righthead{Cygnus Loop Panorama}

\begin{document}
\slugcomment{To appear in {\it The Astrophysical Journal Supplement Series}}

\title{Panoramic Views of the Cygnus Loop}
\author{N.~A.~Levenson\altaffilmark{1} and James~R.~Graham}
\affil{Department of Astronomy, University of California, Berkeley, CA 94720; 
\\
levenson@astro.berkeley.edu, jrg@astro.berkeley.edu}
\altaffiltext{1}{Guest Observer, McDonald Observatory, University of Texas at Austin}
\and
\author{Luke D.~Keller and Matthew~J.~Richter}
\affil{Department of Astronomy, University of Texas, Austin, TX 78712;
\\
keller@astro.as.utexas.edu, richter@astro.as.utexas.edu}
\begin{abstract}
We present a complete  atlas of the Cygnus Loop supernova remnant
in the light of \o3 ($\lambda 5007$), \ha, and 
\s2 ($\lambda\lambda 6717, 6731$).  We include low-resolution ($25\arcsec$)
 global maps and smaller fields at $6\arcsec$  resolution from observations
using the Prime Focus Corrector on the 0.8-m telescope at McDonald 
Observatory.
Despite its shell-like appearance, 
the Cygnus Loop is not a current example of a Sedov-Taylor blast wave.  
Rather,  the optical emission traces interactions
of the supernova blast wave with clumps of gas. 
The surrounding interstellar medium forms the walls of a cavity through which
the blast wave now propagates, including a nearly complete shell in
which non-radiative filaments are detected.
 We identify non-radiative shocks around half the 
perimeter of the Cygnus Loop, and they trace a circle of radius 
$R = 1\fdg 4$ (19 pc) in the spherical cavity walls.  
The Cygnus Loop blast wave
is not breaking out of a dense cloud, but is instead running into
confining walls.
Modification of the shock velocity and gas temperature due to
 interaction of the blast wave with the surrounding medium
introduces errors in estimates of the age of this supernova remnant.
The optical emission of radiative shocks arises only where the blast
wave encounters inhomogeneities in the ambient medium; it is not a
consequence of gradual evolution to a global radiative phase.
Distance measurements that rely on this uniform blast wave
evolution are uncertain, but the radiative shocks can be used as
distance indicators
because of the spherical symmetry of the surrounding medium.
The interstellar medium dominates not only the appearance of the Cygnus
Loop but also the continued evolution of the blast wave.  
If this is a typical example of a supernova remnant, then
global models of the interstellar medium must account for such
significant blast wave deceleration.

\end{abstract}

\keywords{ISM:individual (Cygnus Loop)--shock waves--supernova remnants}

\section{Introduction\label{secintro}}
Supernova remnants greatly determine the large-scale structure of
the interstellar medium.  
The energy of supernova remnants heats and ionizes the
interstellar medium (ISM), and their blast waves govern mass exchange
between various phases of the ISM.  In doing so, supernova remnants
(SNRs) influence subsequent star formation and the recycling of heavy
elements in galaxies.  Global models of the interstellar medium
that include a hot ionized component (\cite{Cox74}; \cite{McK77})
are sensitive to the supernova rate, the persistence of their remnants,
and the sizes they attain.  
A simple calculation of the last of these
assumes that the blast wave expands adiabatically in  a uniform medium
once the blast wave has swept up mass comparable to the mass of the
ejecta.
During this Sedov-Taylor phase, 
the radius of the SNR as a function of $E_{51}$, the initial energy in
units of $10^{51}$ erg, $n_o$, the ambient number density in units of
 ${\rm cm^{-3}}$, and
$t_4$, time in units of $10^4$ yr, is
$R=13 (E_{51}/n_o)^{1/5} t_4^{2/5} {\rm \,pc}$ in a medium where
the mean mass per particle is $2.0 \times 10^{-24} {\rm \, g}$. This phase
will last until radiative losses become important.  The beginning 
of this subsequent phase, marked by the initial loss of pressure behind
the blast wave, occurs at
$t=1.9\times 10^4 E_{51}^{3/14} n_o^{-4/7} {\rm \,yr}$, when
the radius is 
$R=16.2 E_{51}^{2/7} n_o^{-3/7} {\rm \,pc}$ (\cite{Shu87}),
although the radiating shell is not fully formed yet.

We approach these large-scale questions with analysis of complete
images
of a particular supernova remnant, the Cygnus Loop, in three optical
emission lines.  
This supernova remnant appears to
be a limb-brightened shell at radio (\cite{Keen73}), 
infrared (\cite{Bra86}), optical (\cite{Fes82}), and X-ray (\cite{Ku84}; 
\cite{Lev97})
 energies, which at first glance suggests that it is presently in
the transition to the radiative stage.
  The Cygnus Loop has the advantages of being nearby, bright, and relatively 
unobscured by dust.  This allows us to examine in detail
the evolution of various portions of the shock front and to determine
physical parameters, such as shock velocity and local ambient density, 
as they vary throughout the remnant.  
Despite its appearance, the Cygnus Loop is not a current example of blast 
wave propagation in a uniform medium at any stage.  Instead, its evolution 
is governed by the inhomogeneous interstellar medium, which we map using 
the shock as a probe.  

Many of the features we discuss have been noted by others.  Oort (1946)
\nocite{Oort46}
first suggested that the Cygnus Loop is an expanding supernova shell.
Spectroscopy of radiative shocks in selected locations (e.g., \cite{Mil74},
\cite{Ray80a}, and Fesen et al. 1982\nocite{Fes82}) 
combined with theoretical models of these
shocks (e.g., \cite{Cox72}, \cite{Dop77}, \cite{Ray79}, and \cite{Shu79})
 has been used to derive the physical conditions of the observed shocks.
  We utilize the radial velocity
measures of Minkowski (1958)\nocite{Min58}, 
Kirshner \&\ Taylor (1976)\nocite{Kir76}, Greidanus \&\ Strom (1991)\nocite{Gre91},
 and Shull \&\ Hippelein (1991)\nocite{Shu91} to 
discern some of the three-dimensional structure that is ambiguous from
the data we present.
Many non-radiative or Balmer-dominated shocks in the Cygnus Loop have been
identified (e.g., \cite{Kir76}, \cite{Ray80b}, \cite{Tref81}, 
Fesen et al. 1982\nocite{Fes82}, \cite{Fes92}, and \cite{Han92}).
Our observations
qualitatively match these, and we rely on 
these works and others (\cite{Ray83}; \cite{Long92}; \cite{Hes94})  
for quantitative measures of
parameters such as shock velocity and preshock density.
McCray \&\ Snow (1979)\nocite{McC79} and 
Charles, Kahn, \&\ McKee (1985)\nocite{Cha85} have suggested that the Cygnus 
Loop is the result of a cavity explosion, and we adapt this
global model to interpret the surrounding interstellar medium, as well.

This paper is a companion to the soft X-ray survey presently in progress with
the {\it ROSAT} High Resolution Imager (\cite{Gra96}; \cite{Lev97}).
With these two surveys, we examine the Cygnus Loop as a whole,
not restricting our investigation only to 
those regions that are exceptionally bright or that appear to be
particularly interesting.  
We hope to understand both the global
processes and the specific variations that are responsible for the
emission we detect.  The
X-rays probe hot (temperature $T\sim 10^6$ K) gas that shocks with velocities
$v_s \sim 400 \kms$ heat.  The optical emission is expected from
slower shocks ($v_s \lesssim 200 \kms$) in which  the post-shock region
cools to temperatures $T\sim 10^4$ K, yet the most prominent regions at
optical wavelengths are also bright in X-rays.  
McKee \&\ Cowie (1975) \nocite{McK75}
suggested that the broad correlation of X-ray and
optical emission is the result of a blast wave propagating in an 
inhomogeneous medium.  In this scenario, the
shock is significantly decelerated in dense clumps of gas, while portions of it
proceed unimpeded through the lower-density intercloud medium.  
We apply the principles of this basic cloud--blast-wave interaction to
a range of locations in the Cygnus Loop.
In particular, we refine the cavity model introduced in Levenson et al. 
(1997)\nocite{Lev97}, using these optical data to constrain the current
ISM in the vicinity of the Cygnus Loop and to determine how the
stellar progenitor modified it in the past.

We present the observations in \S 2. We describe them in detail,
noting individual regions of interest, and we use these data to measure
the physical conditions of the blast wave and the ambient medium in 
particular locations in \S 3.  The purpose of the detailed examination
is to combine the results in a complete map of the surrounding ISM.  We
present this three-dimensional model while 
providing a coherent explanation of the history that
accounts for it in \S 4.
We predict the future of this SNR and relate its fate 
to more general theories of
supernova modification of the interstellar medium in \S5 and summarize
our conclusions in \S 6.

\section{Observations}
We obtained narrow-band images of the Cygnus Loop
in 1995 August and September  and  1995 September 
on the 0.8-m telescope at McDonald Observatory with the 
Prime Focus Corrector (PFC; \cite{Cla92}).
    The PFC system is a 5-element catadioptric corrector located at the F/3
prime focus of the telescope and projects a flat focal plane onto a
Loral-Fairchild $2048 \times 2048$, $15\micron$-pixel charge-coupled device.  
The camera and corrector system has a
$0\farcs3$ spot size in the V-band and projects each pixel to $1\farcs35$
 on the sky, producing a $46\arcmin \times 46\arcmin$  field of view.
The interference filters each have a rectangular transmission profile
and full width at half maximum $FWHM = 40$\AA\ 
in the converging beam of the PFC.   The filter centers in the F/3 beam
are 5007\AA, 6723\AA, and 6563\AA\ 
to detect line emission from
\o3 ($\lambda5007$), \s2 ($\lambda\lambda6717, 6731$), and \ha+[\ion{N}{2}],
which we refer to throughout as \ha.
  Each individual
integration of approximately $600$\ s duration was bias-subtracted, 
flat-fielded, and scaled by its median sky value before being combined 
with other images.
Every region of the Cygnus Loop was observed at least twice through each
filter in order to remove cosmic rays and
exclude bad regions of the detector, and most regions were observed three 
times in \ha.  We obtained a total of 62 observations in \o3, 89  in \ha, 
and 54 and \s2.

Over the large spatial scale of the Cygnus Loop ($3^\circ \times 3\fdg 5$), 
each observation is a distinct
projection of the sky onto the plane of the detector, so images must
be remapped to a common projection before  being co-added.
An astrometric solution for each integration was calculated based on
the {\it Hubble Space Telescope} Guide Star Catalog.  Alignment 
does not rely on 
overlapping observations of a particular region but depends 
instead on the astrometry of individual integrations.  The final images
are the result of pixel-by-pixel averaging in the
combined image plane using  nearest-neighbor matching and rejecting
bad regions of the detector.  
The global maps of the Cygnus Loop with $25\arcsec$ pixels
are shown  with linear
scaling in \ha, \o3, and \s2, and in a false-color combination with logarithmic
scaling
(Figure \ref{figwhole}).  While the observations at different 
wavelengths resemble
each other broadly, their variation is apparent in the combined image.
In this display, blends of \ha\ and \s2\ alone are yellow, 
\o3\ and \s2\  together appear as magenta, and \ha\ and \o3\ 
combine to make cyan.  Only the regions that are displayed as white
exhibit strong emission at all three wavelengths.
In addition to the global maps, we present 5 smaller fields, each  $95\arcmin$
across, at $6\arcsec$ resolution (Figs. \ref{figsubv}--\ref{figsubz}).
Although we first describe the current physical conditions in 
specific regions of interest, our ultimate goal is to combine the understanding
of the particular sites into a complete account of the surrounding ISM,
its modification by the progenitor star, and the passage of the blast
wave through it.

\section{Emission Morphology and Structure of the ISM} 

\subsection{Observed Morphologies}
The radiative shocks that are responsible for most of the optical emission
from the Cygnus Loop result in two characteristic morphologies: filaments
and diffuse emission.  
Figure \ref{figann} identifies examples of these features and other
structures to which we will refer by name.
The morphological distinction is the result of different viewing 
geometry.  As the ``wavy sheet'' model (Hester et al. 1983\nocite{Hes83}; 
\cite{Hes87}) describes, 
bright, sharp filaments are the result of long lines of sight 
through tangencies of the shock front.  When the shock front is
viewed face-on, the emission appears more diffuse. 

Combined with the X-ray data, these observations provide more specific
information on the evolutionary state of the Cygnus Loop and the distribution 
of inhomogeneities of the ISM around it.
The widespread X-ray emission indicates that the global shock velocity
is several hundred \kms, so the optical emission comes from regions
of denser gas that have decelerated the blast wave.
The correlation of the X-rays with
optically-bright regions make clear that this SNR is not globally
in the gradual transition to the radiative phase that is the result
of blast wave propagation in a uniform medium.  
The blast wave has recently and suddenly decelerated where radiative shocks
are detected.
This rapid, environmentally-imposed evolution is distinct from the
simple model prediction for incremental change in a homogeneous medium.

These complete data sets demonstrate that many characteristics  
that have been investigated in small regions of the Cygnus Loop
are widespread, and we adapt the physical explanations of these limited
regions in previous work to understand the Cygnus Loop globally. 
For example, the bright emission from radiative shocks 
is found to be stratified at select observed locations
(\cite{Fes82}), and we show this stratification to be typical of
all prominent radiative shocks.  
The \o3\ extends toward the projected exterior,
and the \ha\ and \s2\ are located toward the
projected interior.  The \s2 is farthest to the interior and is particularly
diffuse.
The stratification is a function of the column density that the
shock has swept up.
The incomplete \o3\ emission is concentrated behind shocks that have 
encountered relatively little gas, while where the shock lags because
it has swept up a greater column of gas,
\ha\ and then \s2\ are prominent.  The strong \o3\ emission
requires that the shock has progressed through a column density
$N \approx 3\times 10^{17} {\rm \ cm^{-2}}$, while the complete recombination
zone develops after the swept-up column 
$N \gtrsim 10^{18} {\rm \ cm^{-2}}$ for $v_s < 150 \kms$ (\cite{Ray88}).
The apparent thickness
of these zones and the extensive regions of bright emission in a 
combination of all the observed lines is a consequence of projection
effects.  The curved surface of the blast wave encounters the cloud
at different projected radii, so some lines
of sight intersect multiple stages of shock evolution.

Around most of the perimeter of the Cygnus Loop, filaments are
detected in \ha\ alone.
These Balmer-dominated filaments
are the result of non-radiative shocks (\cite{Che78}; \cite{Che80}).  
Unlike the more common radiative shocks,
the \ha\ emission here does not come from the extended cooling zone behind the
shock.  Instead, it is the result of collisional  
ionization before excitation in the immediate post-shock region, or
the downward cascade toward the ground state following charge exchange.
In either case, neutral atoms are required in the post-shock region.
Because these do not survive long, the emission originates at the
shock front.  Thus, we use the Balmer filaments to
trace the shock front directly.
This \ha\ emission is
intrinsically weak, so it is most obvious when the line of sight through
the  shock front is long.  This is the case at the projected
edge of the SNR, where the line of sight is tangent to the spherical
blast wave.

\subsection{Projected Edges\label{subsecedge}}
The projected edges of the Cygnus Loop are the simplest regions to understand
because the geometric complications are minimized there.  
The observed structures may be slightly---but not significantly---in the
foreground or background and still appear near the projected edge.  
The prominent regions, NGC 6992 at the
northeast ($\alpha=\hms{20}{56}{24},\delta=31^\circ 43\arcmin$; 
Fig. \ref{figsubv})
and NGC 6960 at the west ($\alpha=\hms{20}{45}{42},\delta=30^\circ 43\arcmin$;
Fig. \ref{figsubw}),
are examples of interactions of the
blast wave with large clouds
(\cite{HesCox}; Hester et al. 1994\nocite{Hes94}; \cite{Gra95}; \cite{Lev96}),
and the corresponding X-ray enhancement arises from reflected shocks that
propagate through the hot, shocked SNR interior.

The observed length scales of optical and X-ray emission from NGC 6992 
and NGC 6960 reveal the sizes of these large clouds.
At the 770 pc distance of the Cygnus Loop, the coherent networks
of filaments over $20\arcmin$ imply that
the  clouds have lengths of about $10^{19}$ cm. 
Balmer filaments, high-resolution X-ray observations 
or both constrain
the location of clouds  along the line of sight.  X-ray emission extends beyond
both of these bright optical edges, 
where some portion of the three-dimensional
blast wave is not yet impeded by the clouds.  Thus, the clouds do not extend
over the complete edge of the blast wave along the line of sight.

In contrast, the southeast knot at
$\alpha=\hms{20}{56}{20},\delta=30^\circ 25\arcmin$
(Fesen et al. 1992\nocite{Fes92}; \cite{Gra95})
 is an example of a cloud that is
clearly extended along the line of sight but is not necessarily
large across the plane of the sky.  The nearly-circular edge of the X-ray 
SNR and the Balmer filaments are
 distinctly concave at the southeast knot (Fig. \ref{figsubx}a), and there is no
X-ray emission to the exterior.  Thus, this cloud must be at least
11 pc long along the line of sight in order to impede the entire
projected edge of the blast wave at this location.  Again, the
presence of the complete optical cooling structure indicates that the shock
has progressed through a column $N \gtrsim 10^{18} {\rm cm^{-2}}$
here.

The breakout to the south, centered on 
$\alpha=\hm{20}{50},\delta=29^\circ 10\arcmin$ and extending approximately
$1^\circ$ across
 (Fig. \ref{figsubz}), is the most significant 
departure from 
circularity in the remnant.  The greater advance of the blast wave
in the breakout compared with the circular loop north of 
$\delta=29^\circ40\arcmin$
indicates that the ambient density is lower in this southern region.
Balmer-dominated filaments surround much of the breakout, where the
blast wave has now encountered atomic gas.  There are sections of
incomplete shocks to the east and west (e.g.,
$\alpha=\hm{20}{53},\delta=29^\circ 20\arcmin$ and
$\alpha=\hm{20}{48},\delta=29^\circ 15\arcmin$)
and fully radiative shocks
at the east
($\alpha=\hm{20}{52},\delta=29^\circ 15\arcmin$).  
Emission is absent across the face of this area.
In the plane of the sky, the breakout is not very extended.  Either
the blast wave has only recently entered this low-density region, so
it has not had enough time to advance significantly, or the extent
of the low-density material is limited, so the blast wave no longer
continues to progress rapidly.
The stratification of emission around the edge of the breakout is apparent on
smaller scales than those observed elsewhere in the Cygnus Loop, and 
the breakout exhibits smaller regions of common overlap.  
In the northeast, for comparison, the distinct regions of \ha\ and \o3\ are
typically $3\arcmin$ wide, and they are observed together 
over $10\arcmin$ scales,
whereas in the breakout, the exclusive and common zones are both
typically $1\arcmin$ wide.
These
characteristics are consequences of the more limited scale of
interactions with clouds in the breakout.  There is no evidence for indentation
or blast wave progress beyond the optical-emitting regions, which could
be due to either intrinsically small clouds or early stages of encounter.
In either case, at least some dense material surrounds the SNR
at its southern extremity.

The exterior of the southwest quadrant exhibits many filaments that are
nearly aligned with one another (Fig. \ref{figsuby}).  
Some of these are part of the radiative shock structure
related to the southern end of NGC 6960, while farther away from the 
center, these are non-radiative and incomplete shocks (Fesen et al. 1992\nocite{Fes92}).
Along some of the filaments, the transition from
non-radiative to incomplete shock structure is clear
 where \o3\ emission arises.  The multiple filaments are easily understood
as tangencies of a wavy sheet.   These filaments are shorter than those
at the north and east of the Cygnus Loop.  This indicates that the
ISM is clumpier here, which would distort the blast wave in many places 
and  account for the number of filaments that
we observe.  Shock diffraction through low-density 
($\delta \rho/ \rho \gtrsim 1$) regions 
 may be responsible for the large
curvature of some of these filaments, as Fesen et al. (1992) suggest,
although multiple interfering shock fronts are not necessary to produce the different
filaments.  They arise from multiple projections of the same shock front.

\subsubsection{Balmer Filaments and X-ray Emission}
The Balmer-dominated filaments around the periphery of the Cygnus Loop 
indicate that the fraction of atomic gas surrounding the SNR is high.
The extreme circularity of these filaments requires
that the blast wave remain nearly spherical with little deformation at
the projected edge.  
With the exclusion of the breakout region, 
we fit them with a circle of radius $R=1\fdg 4$, which is equivalent to
19 pc at a distance of 770 pc.
In Figure \ref{figcirc}, the model circle is drawn on the soft X-ray image 
obtained with the {\it ROSAT} High Resolution Imager (\cite{Lev97}).
We calculate the correlation of radial variation
as a function of angular separation and the corresponding Fourier
coefficients for the Balmer filaments.  
Excluding the breakout, the amplitude of the
$m = 2$ mode is less than 0.02, which implies that the surroundings
of the Cygnus Loop are homogeneous on large scales.

The individual filaments
tend to be very long, typically extending over more than $40\arcmin$.
The continuity and smoothness of the Balmer
filaments indicate the uniformity of the medium in which these shocks
propagate.  For constant ram pressure $\rho v_s^2$ behind the shock, 
variations in density and velocity are related 
$\delta \rho / \rho = -2 \delta v_s / v_s$.  In terms of the radius, $R$, and
time, $t$, $v=\eta R/t$, where $\eta$ is of order unity; for free
expansion $\eta =1$, and for an adiabatic supernova remnant in a uniform
medium, $\eta = 2/5$.
Thus, $\delta \rho / \rho = (-2/\eta) (\delta R/ R)(t/\delta t)$.
The observed radial variations of a particular filament  
are  $\delta R/ R \sim 0.01$.
An approximate timescale of these shocks is 1000 yr, and the age of the
Cygnus Loop is about 14,000 yr (cf. \S \ref{secage}), so
the overall
density variations in the present medium of the Balmer-dominated filaments
are  of order $\delta \rho / \rho \sim 0.3  \eta$.  We expect 
$ 2/5 < \eta < 1 $, so $0.1 \lesssim \delta \rho / \rho \lesssim 0.3$.

The Balmer-dominated filaments define the current location of
the blast wave and mark the presence of neutral material.  
Detailed studies of particular locations at the northeast have used these
non-radiative shocks as density probes 
(\cite{Ray83}; \cite{Long92}; Hester et al. 1994\nocite{Hes94}) 
and derive densities $n \sim 1 {\rm \ cm^{-3}}$.
   The total observed 
Balmer line intensity is sensitive to density, shock velocity, and geometric
perspective.  All of the filaments around the periphery share the same
 edge-on geometry.
Because of the remnant's circularity, it is reasonable to assume 
that the shock velocity is nearly
constant around the extreme projected edge except at the
breakout.  The  filaments that have been studied in detail to determine
shock velocity and ambient density are the brightest ones,
 however, so these regions are most likely in the
transition to incomplete radiative emission, and these densities are
expected to be higher than elsewhere.
  Thus, we conclude that where typical
 non-radiative filaments are observed around the periphery,
 the  density is constrained $n \lesssim 1 {\rm \, cm^{-3}}$, and most likely
$n \sim 0.1 {\rm \ cm^{-3}}$.  

We roughly calculate the \ha\ surface brightness of a non-radiative shock in
the Cygnus Loop.  We examine a portion of the blast wave 
in the southeast, near  $\alpha=\hm{20}{56},
\delta=30^\circ 00\arcmin$.  This is not an exceptionally bright filament,
and there is no obvious interaction with a large cloud here.  
The whole filament extends over more than $20\arcmin$, so the
blast wave edge is smooth here.  We estimate a surface brightness of 
$4.2\times 10^{-6} {\rm \,erg\,s^{-1}\,cm^{-2}\,sr^{-1}}$ from this
filament. Assuming that this line of sight passes through two folds of the
spherical blast wave and the ISM does not distort it, the intensity
of a single shock surface is 
$2.1\times 10^{-6} {\rm \,erg\,s^{-1}\,cm^{-2}\,sr^{-1}}$.  
The line intensity, $I$, depends on the energy of the transition, $h\nu$,
and is given by
$$I = {{h \nu}\over{4\pi}}n_H v_s{q_{ex}\over{q_i}} {\rm \,erg\,s^{-1}\,cm^{-2}\,sr^{-1}},$$ 
where $n_H$ is the preshock neutral density and  
the ratio of excitation and ionization rates  $q_{ex}/q_i \approx 0.2$
for \ha\ (\cite{Ray91}).  If we reasonably constrain the shock velocity
$200 < v_s < 400 \kms$, these observations limit the density
$0.3 > n_H > 0.15 {\rm \,cm^{-3}}$ at the location of the non-radiative shock.
The lower velocity limit is based on observations of Balmer filaments
(\cite{Ray83}; \cite{Long92}; Hester et al. 1994\nocite{Hes94}), which are
likely to represent the slowest non-radiative shocks, as noted above, 
and the upper bound assumes that the forward shock does not heat the 
X-ray-emitting medium to temperatures greater than $2 \times 10^6 {\rm \, K}$.

The Balmer filaments are related to adjacent X-ray emission:
the optical filaments form the exterior 
boundary of low surface brightness, limb-brightened X-rays (\cite{Lev97}).
This characteristic morphology reveals where the blast wave is 
decelerated in the cavity walls.
We combine the density determined from the Balmer observations with
X-ray temperature measurements to calculate the 
unperturbed blast wave velocity and the original density of the cavity,
and we compare this with the cavity density based on the observed X-ray
surface brightness of the interior of the SNR.
A bimodal temperature distribution has been detected in the Cygnus Loop,
with peaks at $1.5 \times 10^6$ and $5 \times 10^6$ K in EXOSAT
observations with the channel multiplier array (sensitive over 0.05--2 keV) 
and the medium energy experiment (sensitive over 1--50 keV; \cite{Bal89}).
We suggest that the higher temperature component is due to the reflected shock
and that previous low single-temperature measures are contaminated by the
slow, forward shock in the cavity walls.
We model these simply as a reflected shock
and ignore non-equilibrium effects, which are probably important.
The physical parameters temperature, pressure, and density, are related
by the equations of conservation of mass, momentum, and energy flux 
across shock fronts.  
Assuming the temperature of $1.5 \times 10^6 {\rm \,K}$ corresponds to the
decelerated shock in the cavity wall and $T=5 \times 10^6 {\rm \,K}$
corresponds to the reflected shock, the current blast wave velocity
is around $300 \kms$, and the original, unperturbed blast wave
velocity was around $500 \kms$.

The reflected shock model also predicts
that the original cavity density, $n_c$, is related to
the shell density $n_s/n_c = 5$.
Combined with the surface brightness measurement
of the Balmer filaments, this yields an atomic shell density 
$n_s = 0.2 {\rm \,cm^{-3}}$ and a cavity density
$n_c = 0.04{\rm \,cm^{-3}}$ if the shell is entirely neutral. 
As described below (\S\ref{subsecmodel}), 
the shell is partially ionized, so 
the original cavity density was $n_c = 0.08 {\rm \,cm^{-3}}$.
In addition to neglect of non-equilibrium effects, other sources 
of error include confusion of the reflected
shock and the unperturbed shock in the cavity,
uncertainty of the derived X-ray temperature, and uncertainty of the
Balmer intensity of a single shock surface derived from observation through
the spherical edge of the blast wave.

The predicted density is consistent with that derived from X-ray observations
of the projected interior.  In a featureless region
of the southeast near $\alpha=\hm{20}{55},\delta=30^\circ 12\arcmin$,
the median HRI surface brightness is 
$0.01 {\rm \, counts \, s^{-1} \, arcmin^{-2}}$.  We compare this with a 
model of a Raymond-Smith plasma at $T_e = 3.5 \times 10^6$, correcting for
absorbing column of $N_H = 6\times 10^{20} {\rm \, cm^{-2}}$.  Assuming 
the emitting region is 22 pc deep (the predicted line of sight through
the SNR at this radius), the observed count rate corresponds to
a density  $n_e = 0.2 {\rm \, cm^{-3}}$.  
This is the density of the shocked material;
the original cavity density would have been $n_c = 0.05 {\rm \, cm^{-3}}$.

\subsubsection{Photoionization}

One unusual feature at the east is a region of very smooth
emission adjacent to the bright structure of NGC 6995 at
$\alpha=\hms{20}{57}{30},\delta=31^\circ 00\arcmin$.  
This area is obvious in
\ha\ and detectable in \s2 but absent in \o3.  There are no X-rays
here (\cite{Lev97}), so a fast 
shock wave has not passed through.  The X-ray map shows
a clear indentation in this region, where dense gas has significantly
impeded the blast wave over large scales, similar to the
southeast cloud.  There is no other direct evidence,
such as a Balmer filament, that the blast wave is presently located
toward the exterior of this region.  We propose that this gas 
is photoionized by a shock precursor, and it has not been shocked.  

Radiation from the shock-heated gas of the interior of
the Cygnus Loop serves as the ionization source.  
Typically, when the shock velocity
$v_s \gtrsim 100 {\rm \ km \ s^{-1}}$, much of the subsequent radiation is at
UV wavelengths (\cite{Ray79}; \cite{Shu79}; \cite{Dop84}), which
can ionize the surrounding gas, both
ahead of and behind the shock front.   The hardness
of the ionizing spectrum increases with shock velocity, and 
the resulting emission from photoionized
gas is a function of shock velocity and preshock density 
(\cite{Sut93}; \cite{Dop96}; \cite{Mor96}).  The absence of \o3\ 
in the ionized precursor indicates that the shock  has intermediate velocity
($v_s \lesssim 200 {\rm \ km \ s^{-1}}$), which is consistent
with  UV spectroscopy of 
Blair et al. (1991)\nocite{Bla91} who find $v_s \sim 170 \kms$ is typical
in this region.  The bright \ha\
emission ahead of the shock front
is probably a consequence of relatively high density;
the Balmer luminosity of the precursor scales linearly with density 
in the steady case
(\cite{Ray79}; \cite{Shu79}; \cite{Dop96}).  A second important factor is that
the shock is fully radiative, having been decelerated in the dense cloud,
so the shock front does not move through the ionized region before
the ions have time to radiate (cf. \cite{Ray88}).
Finally, the column density ahead of the shock must be great 
for the precursor emission to be observable.

The images do allow us to estimate
crudely the density of the photoionized material.  We assume that the
length of the photoionized region along the line of sight is equal to 
its extent north-south in the plane of the sky and use the observed 
surface brightness to predict a density $n\sim 60 {\rm \, cm^{-3}}$.
Spectral data would allow a more exact determination of the characteristics 
of the emitting region and quantitative measurements of the photoionizing
source.

The blast wave history causes the ambient medium to be  photoionized 
here and not elsewhere.  The penetrating X-rays of very fast shocks 
($v_s \gtrsim 400 \kms$) do not efficiently
ionize the surrounding medium; the optical depth to 200 eV photons
$\tau_{\rm 200 eV} = N/(1.1 \times 10^{20}{\rm \, cm^{-2}})$.
Only the softer UV-dominated spectra
of slower shocks are effective ionizing sources.  In other regions
of the Cygnus Loop, the recent shock velocity has been too great to produce
much photoionization, and there has not been enough time for the
ionizing field of the post-shock cooling and recombination zone to develop.
Thus, most of the surrounding medium has remained sufficiently
neutral to exhibit Balmer-dominated filaments.  The exceptional photoionization
at the eastern edge 
is consistent with the cavity model of this SNR.  The blast wave has not
slowed sufficiently elsewhere for most of the shocked gas to be 
a useful source of ionizing
photons.  Over nearly all of the Cygnus Loop,
the deceleration has been recent, as the blast wave has run into the 
walls of the cavity.  Only in this region has the 
softer spectrum of a slower shock existed long enough to ionize the preshock
medium so that its radiative transitions can be detected.

Photoionized regions have been observed around other SNRs, including
N132D (\cite{Mor95}) and N49 (\cite{Van92}) in the Large Magellanic Cloud,
for example.  
One difference between the Cygnus Loop and the former SNR,
however, is that while in the Cygnus Loop the photoionized precursor is
the exception, detected only in a limited region, in N132D it is
the rule, occurring over the majority of the younger remnant's perimeter. 
Another difference
is that the exceptional photoionization
ahead of the Cygnus Loop blast wave is adjacent to the region that
contains its brightest soft X-ray knots (\cite{Lev97}),
while the photoionization and X-ray surface brightness variations are not
correlated in N132D (\cite{Mor96e}).  These differences
are likely to be a consequence of the different sources of ionizing photons
and the density of the surrounding medium.
First, in N132D, the X-ray--emitting shock can account for
only half of the ionization (\cite{Mor96e}), so another significant
source must also contribute to the ionization in this case.  The
ubiquitous radiative shocks in clouds and O-rich filaments that 
Morse et al. (1996)\nocite{Mor96e} propose as this additional source 
are not preferentially associated with the X-ray emission.
Second, a large molecular cloud extends over the entire southern hemisphere
of this SNR (\cite{Ban97}).  While the present blast wave is not
in the dense core that is observed in CO, the outer regions of the cloud
are expected to be over-dense with respect to the typical ambient ISM.
The higher densities that make the ionizing precursor visible are
therefore characteristic of the majority of the perimeter of N132D.
In contrast, we suggest
that it is the unusual conditions of extreme blast wave deceleration and
related high density that result in both the bright X-ray emission and
the photoionized precursor in this small area 
at the eastern edge of the Cygnus Loop.
These conditions make the Cygnus Loop precursor similar to that of
N49.  This middle-aged SNR is associated with a small molecular cloud
on its eastern limb (\cite{Ban97}).  
The blast wave interaction with
the cloud results in local optical emission and X-ray enhancement, 
blast wave deceleration observed in a range of shock velocities
(\cite{Van92}), and a prominent photoionized precursor on the
eastern limb of N49.

\subsection{Projected Interior} 
The interior of the Cygnus Loop contains diffuse emission and filaments.
Both morphologies share some characteristics, such as
exclusive \o3\ emission 
(e.g., $\alpha =\hm{20}{48}, \delta=30^\circ 30\arcmin$ in diffuse regions and
$\alpha =\hm{20}{52}, \delta=30^\circ 30\arcmin$ in filaments),
and blends of \ha\ and \s2 
(e.g., $\alpha =\hm{20}{50}, \delta=29^\circ 45\arcmin$ and
$\alpha =\hm{20}{50}, \delta=29^\circ 45\arcmin$).
A combination of emission in all 
three lines is observed only in diffuse emission
(e.g., $\alpha =\hm{20}{52}, \delta=31^\circ 10\arcmin$), and
\ha\ alone is observed only in filaments
(e.g., $\alpha =\hm{20}{50}, \delta=30^\circ 30\arcmin$).

Projection effects are more severe in the interior, but these optical images
and high-resolution  spectroscopy (\cite{Kir76}; \cite{Gre91}; \cite{Shu91})
together help unravel the tangled Cygnus Loop.  In all
cases, the optical emission is the result of the interaction of
the three-dimensional blast wave surface with dense structures
in the ISM.  The emitting regions are truly at the outside of the
SNR, but some appear projected to the interior.  The brighter filaments
tend to have lower absolute radial velocities and narrower velocity
dispersions than the fainter diffuse
emission (\cite{Kir76}; \cite{Shu91}).
  This is consistent with the model in which filaments are
long lines of sight through the edge of a wavy shock front, and diffuse
emission results when the blast wave is observed nearly face-on.  The shock
velocity is primarily perpendicular to the line of sight in the former
case and along the line of sight in the latter case.  In some instances,
adjacent filamentary and diffuse emission are related, being tangential
and normal views of the same portion of the blast wave.  Elsewhere,
spatially separated shock regions are projected onto the same location
in the observer's view, causing significant confusion.

The extensive diffuse emission at the western interior near
$\alpha=\hm{20}{47}$---$49^{\rm m}$ and  
$\delta=30^\circ 40\arcmin$---$31^\circ 20\arcmin$  
suggests that a large wall of dense gas is present across the entire face
of the SNR (Figs. \ref{figsubw} and \ref{figsuby}).  
The irregularity of the diffuse emission is due to
the clumpiness of the cloud.  This emission is blueshifted with
respect to nearby gas  (\cite{Gre91}; \cite{Shu91})
so it is on the front face.  The bright filaments of the ``carrot''
(at $\alpha =\hm{20}{49}, \delta=31^\circ 40\arcmin$),
 having positive radial velocities, are on the rear face.  

The  filaments at
$\alpha =\hm{20}{49}, \delta=31^\circ 00\arcmin$ (Fig. \ref{figsubw}) 
are distinct on small scales, 
similar to  the edges of the breakout, although here the region of
common overlap is absent.  The origin of the stratification is the
same: the limited physical scale of the interaction with the
intervening cloud, either because the cloud is intrinsically
small or because the encounter is recent and the blast wave has
not progressed far into the cloud.
Along a particular line of sight, only a single coherent section of the 
blast wave
is detected.  Unlike the areas where strong emission is observed at
all wavelengths, here there are not many projected overlapping regions of
various stages of shock evolution.  The filaments that are bright in
\ha\ and \s2\ illustrate an earlier section of the cloud shock than do
the \o3\ filaments.  In this area on the western half of the Cygnus
Loop, there is no planar blast wave that actually moves to the east, however. 
Rather, the {\em interaction} with a cloud is progressing from west to east
as the blast wave surface expands radially.  The intervening cloud
must be oriented along the line of sight to some degree,
as shown in Figure \ref{figcartc}.

Another region of diffuse emission and fainter filaments runs north-south
across the center of the remnant, around
$\alpha=\hm{20}{50}$---$52^{{\rm m}}, 
\delta=30^\circ 20\arcmin$---$50\arcmin$  
 (Figs. \ref{figsubx} and \ref{figsuby}).  The high-resolution spectra
in the vicinity are of the diffuse component alone, which includes very
bright central knots.  These clumps are blueshifted, so these clouds
are on the near face. The westernmost of these filaments, 
at $\alpha=20^{{\rm h}} 50^{{\rm m}} 20^{{\rm s}}$, is detected only in \ha.
The blast wave has run 
into an extended wall of gas, enabling us to 
look through a long edge of the shock front.
These filaments are broader than most of the
other pure Balmer filaments,  which is expected in this scenario.
The broad filaments appear where the blast wave is viewed through a
point of inflection, whereas the sharper, narrow ones are viewed at a
tangency.  For these  non-radiative filaments to be observable,
the wall of gas must be extremely smooth at the present projected
location of the shock front.  The wall of gas is likely to be related
to the source of the rest of the optical emission at the center of the SNR, not
the clouds that are illuminated as the ``carrot,'' although this is uncertain.

\section{History of the Cygnus Loop}
\subsection{Modelling the ISM\label{subsecmodel}}
Except in the single photoionized section, the optical emission traces 
dense shocked regions.  Using some knowledge
of shock evolution and geometry, we map the surrounding ISM.  
The basic principles applicable to these observations are: 
 (1) optical emission due to radiative shocks requires that
the blast wave has swept up a significant column density 
($N \gtrsim 10^{18} {\rm \, cm^{-2}}$), which is likely in regions
of increased density ($n\gtrsim 1 {\rm \, cm^{-3}}$);
(2) sharp filaments
are edge-on views of the blast wave surface, and diffuse emission is the
result of face-on views of this surface;  (3) the optical line emission
from a  radiative shock is progressively dominated
by \o3, \ha, then \s2;  and (4) filaments of \ha\ alone are
the result of non-radiative shocks in material that has a significant neutral
fraction.

With these data, we construct a three-dimensional model of the interstellar 
medium  around the Cygnus Loop .
Over much of its surface, the blast wave
is propagating through large clouds.  Long filaments, detected in
\ha\ around the perimeter and in other lines across the interior, indicate
that a spherical shell of atomic gas surrounds the SNR.  The dense
clumps of gas and the atomic shell together form the walls of a cavity in
which the supernova occurred.  
Slices through the Cygnus Loop parallel to the plane of the sky from
the near side to the far side (Fig. \ref{figcartmap}) show some of the
three-dimensional structure of the ISM before the supernova event.
The center of the Cygnus Loop is at latitude $b = -8^\circ$.
At the rear of the cavity, toward the Galactic plane, is
a large wall of gas that is detected in neutral hydrogen (\cite{DeN75})
and in the IRAS sky survey (\cite{Bra86}).
The cloud that causes the bright northeastern emission is also on the 
far side.  Clumps at the edge of the breakout and 
at the northwest are on the near side.  
Some clouds, including the southeast knot and the molecular gas at the
west, extend along the line of sight and appear in all slices.

The Cygnus Loop is not the result of a supernova explosion in an arbitrary
region of the interstellar medium.  Rather, the blast wave propagates
through a medium that the progenitor star has processed.  
Several characteristics imply a massive progenitor, the sort most
capable of modifying the surrounding ISM, which 
Charles et al. (1985)\nocite{Cha85} and Levenson et al. (1997)\nocite{Lev97}
discuss.  One strong piece of evidence is 
the paradoxical 
morphology of the Cygnus Loop.  At optical and X-ray wavelengths, it
appears to be nearly circular, yet it exhibits tremendous structure on
smaller scales.  The optical data alone could be interpreted as a SNR
in transition from the adiabatic to the radiative phase, but the presence
and correlation of X-rays require that overall the blast wave still has
(or until very recently had) 
a high velocity, $v_s \gtrsim 400 {\rm \ km \ s^{-1}}$.  A model of blast
wave interaction with large clouds accounts for this correlation 
(\cite{HesCox}; \cite{Gra95}; \cite{Lev96}), but such large clouds cannot
be typical of the pre-supernova interstellar medium.  If they were, the
SNR would not appear so circular, and X-ray emission due to clouds
evaporating in the interior would be observed.  Thus, although large
clouds are found around the periphery of the Cygnus Loop, they were not
located within the current blast wave prior to the supernova event.
This structure is expected in the vicinity of an O or B star.
During its main-sequence lifetime, such a massive star homogenizes the
surrounding ISM, creating a uniform region at constant density, 
temperature, and
pressure.  The star's UV emission removes clumps of denser gas within
the H~II region, either destroying the clouds
through photoevaporation or relocating them to the exterior by
the rocket effect (\cite{McK84}).  The result is a homogeneous
spherical region surrounded by dense clouds.  
A star of type earlier than B0  will clear a cavity with a radius of 
about 50 pc (McKee et al. 1984 \nocite{McK84}).  
The Cygnus Loop cavity is smaller, 
so its progenitor must be a later-type star (\cite{Cha85}), yet one
still able to create the photoionized cavity.  Thus, we suggest that a
B0, $M\approx 15M_{\sun}$ star was the progenitor of the Cygnus Loop.

The observer's line of sight through shocked clouds at the
projected edge is preferentially longer than through those that
appear at the projected center.  Photoevaporation of clouds 
progresses radially outward from the central star, so
lines of sight tangent to the cavity are biased to be through the longer
dimensions of surviving clumps.  Thus, optical emission on the projected 
interior is not as bright as on the
projected edge of the SNR, despite their common cause.  The same effect is 
even more pronounced in the X-ray data.  

The atomic shell in which the non-radiative filaments appear
is also a consequence of the progenitor's evolution.
While material is eroded from clouds during
the star's main sequence lifetime, increasing the density of the
H~II region, the radius
of the ionized region decreases as the outer portions recombine, creating
the shell (\cite{Elm76}).  The
atomic gas of this recombined region together with the cloud
remnants on the periphery form the walls of the
cavity (Fig. \ref{figcartmap}).  
The Cygnus Loop blast wave has expanded through the uniform 
cavity and is now hitting the surrounding walls (\cite{Lev97}).

The atomic shell is not actually a sharp density discontinuity.  It is more
likely a gradual transition from the cavity interior to
the maximum shell density.  At the end of the progenitor's main sequence
lifetime, the pressure in the former H~II region  declines as
the gas cools and recombines.  (See \cite{PShu85} for a discussion of this
phenomenon applied to N49.)   
In the extreme case of
a sudden loss of pressure at the boundary, 
a rarefaction wave would move back through the dense shell
resulting in a density profile of the form 
$n \propto (1 + {3r \over {at_{RG}}})^3$ (\cite{Zel66}),
where the distance, $r$, from the original shell edge is constrained
$-3at_{RG} \le r \le 0$  in terms of sound speed, $a$, and
red giant lifetime, $t_{RG}$.  The drop in pressure is not so sudden, however.
Over $t_{RG} \sim 10^6$ for $M\approx 15M_{\sun}$  (\cite{Mae89}),  
about half the atoms in the
former H~II region have recombined.  Thus, the density profile of the
shell edge is likely to be 
intermediate between a sharp discontinuity and the offset power law above.

The Cygnus Loop has been suggested previously to be the result of an
explosion in an incomplete cavity.  Both Falle \&\ Garlick (1982)\nocite{Fal82}
 and Shull \&\ Hippelein (1991)\nocite{Shu91},
for example,
argue for a partial cavity at a density discontinuity, such as the edge
of a molecular cloud.  In these models, only half of the blast wave 
runs into the cavity boundary.  
Consequently, the bright radiative emission 
of NGC 6992 and NGC 6960 and the corresponding X-ray emission 
are physically separated by great distances along the line of sight: 
the former
are the result of  radiative shocks in the cavity wall, and the opposite
hemisphere of the  blast
wave propagates through the low-density medium outside the cloud to cause
the latter.  In contrast, the  model presented here and in 
Levenson et al. (1997)\nocite{Lev97} has a 
completely defined cavity, the atomic shell and dense clouds together
forming its spherical boundary.  The blast wave does not break {\em out} 
of the  edge of the molecular cloud into lower-density material; instead, it 
slows down when it hits a wall of {\em denser} gas.  
Although NGC 6992 and NGC 6960 and the X-rays that appear 
to the exterior of  these radiative regions are 
two slightly different portions of the 
blast wave, they are not significantly separated.  Furthermore, some of the
large clouds are obviously extended along the line of sight, not restricted
to the rear boundary (Fig. \ref{figcartmap}).
These optical and previous X-ray data (\cite{Lev97}) 
demonstrate that the cavity wall is
complete.  All around the projected edges of the Cygnus Loop, the Balmer 
filaments and X-ray enhancements  mark the smooth
component of the cavity wall.
{\em Even the region called the breakout
illustrates where  the blast wave runs into a boundary of dense gas, not 
continued expansion in a low-density medium.}  Only the eastern third of
the projected interior of the Cygnus Loop lacks direct evidence for a
cavity wall, although the smooth atomic shell that is apparent in 
Balmer-dominated
filaments elsewhere would not be detected across the face of the SNR.
Only clumps of gas could be observed, and their distribution depends on the
arrangement that existed before the ISM was processed by the progenitor.
The southeast of the Cygnus Loop is farthest from the galactic plane,
so we expect the original filling factor of clouds to have been lowest there.

Other SNRs, including RCW 86 (\cite{Vin97}) and  N132D (\cite{Mor96e}),
have been identified as cavity explosions.  
[Shull (1985)\nocite{PShu85} 
identifies N49 as a cavity supernova, but Vancura et al. (1992)\nocite{Van92}
argue that the dense ambient material associated with the 
single observed CO cloud accounts for the data better.]
In these cases, interaction of the blast wave with the cavity wall
enhances the observable properties of the SNRs.
RCW 86 shares many features
of the Cygnus Loop, including similar size, 
a limb-brightened (though very incomplete)
X-ray shell (Vink et al. 1997\nocite{Vin97}), and 
Balmer-dominated filaments to the 
exterior of the X-ray emission (\cite{Smi97}).  
N132D is a younger
remnant, and the apparent cavity is smaller.  
The progenitor's wind during its main sequence lifetime
may form the cavity boundary, or
the scale of photoevaporation may have been limited in
this dense environment adjacent to a molecular cloud (\cite{Ban97}).
The bright interior X-ray emission (\cite{Mor96e}) does imply that 
the cavity interior was not uniform. 

\subsection{Age and Distance\label{secage}}
Although the shell-like visual morphology of the Cygnus Loop and its
X-ray limb-brightening broadly suggest that this is an example of a 
supernova remnant in the Sedov-Taylor or adiabatic expansion phase, 
in detail this model is
inconsistent with the data presented here and in 
Levenson et al. (1997)\nocite{Lev97}.  
The observed X-ray and optical limb-brightening
are a consequence of interaction with the inhomogeneous
interstellar medium, not the
result of evolution of the blast wave in a uniform medium.
Assuming that the blast wave has slowed sufficiently 
during its progress
through a uniform medium to produce optical line emission 
(i.e. to  $v_s \lesssim 200 \kms$) yields an
age estimate of about 50,000 yr (\cite{Min58}; \cite{Fal82}).
This is much too great because
the age is inversely proportional to the blast wave velocity, and only
in propagating through dense clouds has the shock been decelerated to 
$v_s \lesssim 200 \kms$.
The Balmer filaments mark the location of the unimpeded, spherical
blast wave, and the 
X-ray emission immediately behind these filaments demonstrates
that the typical shock velocity is $v_s \gtrsim 400 \kms$,
which is still to great for radiative cooling
of the Sedov-Taylor shell to be significant.  This SNR has not made
the global transition to the radiative evolutionary phase.
The X-ray emission alone has been used to determine an age of around
18,000 yr (\cite{Rap74}; \cite{Ku84}), assuming that this is presently
 an adiabatic blast wave expanding in a uniform medium;
$t = 2R/5v_s$, where $R=19{\rm \ pc}$ and $v_s = 400 \kms$.  

The X-ray data provide a more accurate measurement of the age,
but it is still likely to be an upper limit because
these calculations contain uncertainties
in the shock temperature and the velocity that is derived from it.
Only the current unperturbed 
shock velocity is relevant to the age in this model, but any
observation of the SNR includes contributions from a variety of temperatures,
heated by different stages of the blast wave's history.  
Observing the projected edge of the SNR
minimizes confusion due to overlapping projections of different temperature
components, but the interaction of the
blast wave with the inhomogeneous ISM 
greatly alters the shock velocity and 
the corresponding postshock gas temperature.

When the blast wave hits a density discontinuity,
as it does over most of the surface of the Cygnus Loop, 
the continuing shock is suddenly decelerated and a
reflected shock
propagates back toward the interior, further heating the
dense, hot, shocked material.
At a density
contrast of 10, for example, the twice-shocked matter will have a temperature
approximately 1.5 times the singly-shocked material at the same radius,
which leads to an overestimate of the unperturbed shock 
velocity by about $20\%$.  
Both the reflected shock and the forward shock through the dense cloud
enhance the X-ray surface brightness and thus bias 
the mean measured temperature to favor blast-wave--cloud interactions
rather than the undisturbed blast wave.

If the X-ray measurement of 
two temperature components (\cite{Bal89}) due to
the decelerated cloud shock and the reflected shock is correct and
the original velocity of the unperturbed blast wave was closer to
$500 \kms$ (\S\ref{subsecedge}), the corresponding age  $t = 14,000$ yr.
A reliable determination of the age of the Cygnus Loop is required to
determine the initial energy, $E_o$, accurately
(\S\ref{secintro}).  
Adopting the parameters $n_o = 0.08 {\rm \, cm^{-3}}$, $R = 19$ pc, and 
$t = 14,000$ yr, and a partially-ionized cavity interior,
we find $E_o=2\times10^{50} {\rm\, erg}$.

Because the radiative emission shows where the blast wave has run into
clumpy material and decelerated, care is required to use it
to measure the distance to the Cygnus Loop.  Minkowski (1958) \nocite{Min58}
determined the expansion velocity to be $116 \kms$, fitting an ellipse to
  radial velocity measurements as a function of
radius.  This, combined with Hubble's (1937) \nocite{Hub37}
 observed proper motion of filaments of $0.03\arcsec {\rm \ year^{-1}}$ 
yields the frequently-cited distance of 770 pc.  This calculation assumes that
all the optical filaments are part of a common expanding surface.  
In fact, the optical emission reveals different regions of
inhomogeneity in the ISM.  A second complication is the distinct
geometry of the diffuse and filamentary emission on the
face of the SNR.  One is the result of a face-on shock, and the other is
an example of a shock viewed edge-on.  The latter does not have the 
radial velocity structure as a function of projected radius that is
assumed in the model calculation.
A third source of confusion arises because the observed filaments are the
cooling zone behind the shock front, so their apparent motion may
be the result of further development of the postshock region, 
not motion of the same clump of gas.  The cooling zone may be propagating,
without any physical motion of a parcel of gas in the ISM.

Despite these difficulties, however, the radiative shocked gas can
be used to measure distance to the Cygnus Loop or other supernova
remnants when the distribution of clumps in the 
surrounding medium is spherically symmetric.  
In this case, although the emission comes from distinct regions of
the ISM, not a single expanding blast wave surface, the different clouds 
acquire the same velocity when the blast wave passes through them.
This approach is not appropriate where the surrounding medium is
asymmetric, however.   IC 443, for example, is asymmetrical, and
its corresponding distance determinations are highly uncertain
(\cite{Gre84}).

The non-radiative filaments can also be used to determine distance,
as Hester, Danielson, \&\ Raymond (1986) \nocite{Hes86} have done,
finding a distance of about 700 pc.  These
structures are advantageous
because proper motion and spatial velocity can be calculated from the 
same filament.   There is some uncertainty in the shock velocity
because the derived value depends on the amount of 
ion-electron temperature equilibration
(Chevalier et al. 1980\nocite{Che80}).
Measuring proper motion of filaments also introduces some uncertainty.
  All filaments are detected where the observer's line of sight through 
the shock front is long, which may be the
tangent surface of a spherical bubble.  If it is instead the deformed
surface of the blast wave that appears as a filament, however,
the apparent filament will not necessarily persist or
be detectable over the timescales required to obtain proper motion 
measurements.
Long baselines over which to measure proper motion are desirable,
but the oldest observations are sensitive to the brightest filaments.
These bright filaments are exactly the locations
where the shock is likely 
slowed in denser gas, so the present velocity differs from
the spatial velocity that corresponds to the proper motion.
With these caveats in mind, we adopt a distance of 770 pc to the Cygnus Loop.

\section{Consequences} 

The Cygnus Loop blast wave is now in the walls of the cavity, the 
boundary between the low-density former H~II region and the ambient
ISM.  Blast wave propagation in the atomic shell will be a long-lived
phenomenon.  The smooth material of the shell recombined as
clumps were photoevaporated in the interior.  We crudely calculate the
thickness of the shell assuming a constant number of ions in the H~II region
in both the initial Stromgren sphere, in which only the less-dense intercloud
material was ionized, and in the final Stromgren sphere of radius $R_f$, 
in which all material, having mean density
$n_m$, is ionized.  The typical, unprocessed ISM determines the
initial conditions; we assume the intercloud density
$n_{ic}=0.1 {\rm \, cm^{-3}}$, and a volume filling factor of 0.03 of 
clouds of density $n_{cl}=1 {\rm \, cm^{-3}}$, so 
$n_{m}=0.13 {\rm \, cm^{-3}}$.  Conserving ion number and neglecting
increased recombination as clouds are evaporated, we obtain for
the shell thickness
$$R_{sh}=R_f ((n_m/n_{ic})^{1/3} - 1),$$ so using the observed radius
$R_f = 19{\rm \,pc}$ yields $R_{sh}=1.5 {\rm \, pc}$.  A blast wave of velocity
$v_s= 300\kms$ will propagate through the shell for 5000 years.  

Eventually the blast wave will propagate through the denser
medium of the cavity walls and clouds everywhere. 
It will rapidly decelerate, not
only in the select locations where it has already
encountered dense clumps that protrude into the cavity.
  Some of the surrounding material consists of
large dense clumps like the ones observed presently in the very
extended regions of radiative shocks.  These, too, will light up
and will be detectable at optical wavelengths.  The southeast cloud
 may be an example of the earliest stages of interaction with a
large cloud, and it may later resemble the northeast region (\cite{Gra95}).  
We also expect the entire
northern limb, which in the optical regime is now detected 
primarily in bright Balmer filaments, to turn into an extended
radiative region.  The incomplete shocks along the northern limb are the
early stages of interaction with a larger structure that
 is perhaps directly associated with the
bright northeastern region.  The entire western face of the 
Cygnus Loop will soon be bright when the blast wave has
fully encountered the wall of gas that we have noted there.
If the long filaments across the center of the SNR are the
result of another wall of gas, the middle will also light up with
optical emission in several hundred years.

Large clouds of gas will significantly affect blast wave evolution.
The shock will no longer be adiabatic when the cooling timescale
becomes comparable with the dynamical timescale.  In large clouds, 
where density $n_{cl}=1 {\rm \, cm^{-3}}$, radius $R_{cl}= 5 {\rm \, pc}$, 
and in which the shock velocity $v_{cl}=200\kms$, 
the dynamical time $R_{cl}/v_{cl} \approx 2\times 10^4 {\rm \, yr}$.
In these clouds, the
cooling time $3kT/n_{cl}\Lambda(T)\approx 4 \times 10^4 {\rm \, yr}$, 
in terms of the cooling function, $\Lambda$, having units
${\rm erg\, cm^3\, s^{-1}}$.
If the covering fraction of clumps in the surrounding medium is sufficiently
great, they will force the Cygnus Loop into the radiative phase,
and energy losses will further decelerate the blast wave.  The
time and radius at which this occurs 
 are distinct from self-similar calculations, such as those of
McKee \&\ Ostriker (1977)\nocite{McK77} and
Shull \&\ Draine (1987)\nocite{Shu87}
in which the gradual transition to
the radiative phase happens as the blast wave propagates through
a uniform medium.  

We use the pressure-driven snowplow model (\cite{McK77}) 
to set an upper limit on the final size of the SNR.  In this model,
work of the blast wave pushing into the ambient material and
radiative losses of the hot interior are taken into account.
In the Cygnus Loop, the radiative losses of the outer
shell also will be significant, slowing
the blast wave and making the final radius smaller.
During this phase,  $PV^\gamma$ is constant, where
$P$ is the pressure inside the SNR, $V$ is its volume, and $\gamma$
is the ratio of specific heats.
For $\gamma = 5/3$, this phase ends 
when $R = R_{PDS}(P_{PDS}/P_{ISM})^{1/5}$, in terms of
the external pressure of the interstellar medium, $P_{ISM}$, 
and the radius and  pressure inside the supernova remnant
at the onset of the pressure-driven snowplow stage, $R_{PDS}$ and
$P_{PDS}$, respectively.  At this final time, the pressure inside the
supernova remnant is equal to the ambient pressure.  
For the Cygnus Loop, we take 
$P_{ISM} = 10^{-12} {\rm \, dyne\, cm^{-2}}$, $R_{PDS}=20 {\rm \, pc}$, 
and $P_{PDS} = 4\times 10^{-9} {\rm \, dyne\, cm^{-2}}$, which places
an upper limit on the final radius $R < 105 {\rm \, pc}$.

Assuming that blast waves propagate through uniform media over
great distances yields misleading conclusions about the  size
of SNRs and how they distribute energy through the ISM.
Global models of the interstellar medium that include the hot contribution
of SNRs (\cite{Cox74}; \cite{McK77}) are sensitive to the supernova rate,
the final sizes of supernova remnants, and their persistence as hot
bubbles.  These factors determine whether the dominant component of
the ISM is hot and in which smaller regions of warm and cold gas are embedded,
or whether the warm and cold components form the substrate in which the
distinct hot bubbles of SNRs are located.  For the former to occur, the onset
of the radiative phase of the SNR must be postponed until 
$R > 100 {\rm \, pc}$, independent of its prior evolution (\cite{Cox86}).
(This calculation assumes that the supernova rate is 1 per 30 yr in a disk
of radius 15 kpc, yet the conclusion holds for a range of reasonable
supernova rates.)  
Although the cavity of the Cygnus Loop is small because its
progenitor is one of the least massive
stars to undergo core collapse,
the initial mass function favors these lower mass (and smaller cavity)
progenitors.
Furthermore, even more massive stars, where cavities of
50 pc radius are expected, will be forced into the radiative phase
and not expand to overlap if the  density of the cavity walls is sufficient.
The cavity walls are indeed likely to be denser on average than the typical ISM
because the walls contain much of the material that was cleared out
of the cavity interior.  
The conditions of the ISM in general, however,
constrain the mass that will be commonly available to form clouds and
cavity walls around pre-supernova stars.  If the hot, low-density medium
is pervasive, cavity walls typically would not be sufficiently massive
to induce the radiative phase early, and SNRs would then normally
attain larger radii before their blast waves decelerate.  In any case,
the typical environments of supernova progenitors must be considered.  
If  dense, molecular material normally surrounds
these stars, the radiative phase will be induced early, even if globally the
ISM has low density.

The cavity structure is  significant also in determination of the mass of
gas that a supernova remnant heats.  The onset of shell formation 
for a supernova blast wave expanding in a uniform medium occurs
at $R \propto n^{-3/7}$,
so the shocked mass is then
$M \propto n^{-2/7}$.  If the ambient density varies by three orders
of magnitude, the heated mass changes by only a factor of 7, for example.
In contrast, when the supernova environment is a cavity with dense walls,
the initial size of the cavity constrains the mass of the hot interior.
The cavity interior is rarefied, so the heated cavity mass is significantly
lower.  In the Cygnus Loop, assuming a cavity density
$n_c = 0.08 {\rm \, cm^{-3}}$, the heated mass is $20 M_{\sun}$, whereas
in a uniform medium of this density, the heated mass would be greater than
$400 M_{\sun}$.

\section{Conclusions}
This optical emission line 
atlas of the Cygnus Loop supernova remnant
provides the information to render a portrait of
the surrounding ISM.  The ambient medium,
which the massive stellar progenitor shaped as
it evolved, consists of many large, dense regions.
These will significantly affect the subsequent development of the blast wave.
The blast wave will decelerate, and it will no longer have sufficient
velocity to excite
the material through which it propagates to X-ray-emitting temperatures.

Although this is a study of a particular object, it has
far-reaching consequences for the ISM as a whole.  Supernova remnants
provide the energy to heat the ISM, affect the velocity dispersion
of interstellar clouds, and set the stage for future generations of
star formation.  Thus, any global model of the gas in the Galaxy
critically depends on the evolution of SNR blast waves in their inhomogeneous
environments, of which this example is typical.  
While optical emission-line characteristics identify supernova remnants,
they may preferentially select  those whose evolution the extant ISM 
determines.  Other SNRs need to be examined over a broad range of
energies in the same careful way to draw more certain conclusions
about their net effect on the interstellar medium.

\acknowledgements
This research was supported in part by the Packard Foundation.

\newpage
\figcaption{Complete narrow-band maps of the Cygnus Loop obtained using the
PFC at McDonald Observatory.  Each individual integration covered a
 $46\arcmin \times 46\arcmin$ field using $1\farcs 35$ pixels.
 The mapping procedure used to produce these mosaic images, which
have $25\arcsec$ pixels, is described in the text.  
{\it (a)} The H$\alpha$ map, a combination of 89 individual
observations, is presented with linear scaling. 
{\it (b)} The [O III] map includes 62 observations and is scaled linearly.
  {\it (c)} The linearly-scaled [S II] map is the result of 54 integrations.
{\it (d)} A false-color combination of the  three mosaics uses
[S II] in red, H$\alpha$ in green, and [O III] in blue.  Blends of 
H$\alpha$ and [S II]
alone are yellow, [O III] and [S II] together appear as magenta, and
H$\alpha$ and [O III] combine to make cyan. The individual
frames were median-filtered and scaled logarithmically before
being combined into the false-color image.
\label{figwhole}}

\figcaption{These mosaic images of portions of the Cygnus 
Loop are $95\arcmin$ across and have $6\arcsec$ pixels. 
{\it (a)} The northeast in \ha, {\it (b)} [O III],
 {\it (c)} [S II], and
 {\it (d)} a false-color combination as in Fig. \ref{figwhole} {\it (d)}.
 \label{figsubv}}

\figcaption{The northwest of the Cygnus Loop, as in Fig. \ref{figsubv}.
\label{figsubw}}

\figcaption{The southeast of the Cygnus Loop, as in Fig. \ref{figsubv}.
\label{figsubx}}

\figcaption{The southwest of the Cygnus Loop, as in Fig. \ref{figsubv}.
\label{figsuby}}

\figcaption{The Cygnus Loop breakout, as in Fig. \ref{figsubv}.
\label{figsubz}}

\figcaption{Many of the features described in the text are noted
on this H$\alpha$ image. \label{figann}}

\figcaption{The true blast wave edge, detected in Balmer-dominated
filaments, is nearly symmetrical.  It is best fit by a circle centered on
$\alpha=\hms{20}{51}{21},\delta=31^\circ 01\arcmin 37\arcsec$ 
with a radius of $1\fdg4$. Here the model circle is 
drawn on the {\it ROSAT} High Resolution Imager 
map of the Cygnus Loop in soft X-rays (Levenson et al. 1997).
Although the X-ray observations are incomplete, having gaps on the
interior near
$\alpha=\hm{20}{49},\delta=30^\circ 00\arcmin$ and 
$\alpha=\hm{20}{48},\delta=31^\circ 30\arcmin$, they
illustrate the blast wave clearly.
\label{figcirc}}

\figcaption{This cartoon illustrates the orientation of a cloud along
the line of sight near the roots of the ``carrot,'' 
at $\alpha =\hm{20}{49}, \delta=31^\circ 00\arcmin$. As the
blast wave expands radially (from $t_1$ through $t_2$ to $t_3$), 
it interacts with the cloud from west
to east in the plane of the sky, and the observer views its apparent progress
along the arrow. \label{figcartc}}

\figcaption{Views of the interstellar medium before the supernova occurred.
The three cartoons represent slices in the plane of the sky on the
far side, through the center, and on the near side of the cavity, with
north at the top and east toward the left.
Features that appear in all three slices extend along the line of sight.
  A large wall of gas associated with the
Galactic plane and the cloud that causes radiative emission at the 
northeast are toward the rear.  The atomic shell is prominent in the
slice through the center of the cavity.  Clouds at the northwest and
around the edge of the breakout are on the near side.  Scoville et al. (1977) 
observed the molecular gas near the
Cygnus Loop.  The extended wall of material at the west is detected in
IRAS sky survey and in HI observations (DeNoyer 1975). 
The other clumps interact with the blast wave and are
evident in the optical observations presented here and in the X-ray
observations of Levenson et al. (1997).  
\label{figcartmap}
}

\end{document}